\documentclass[twocolumn]{aastex62}

\newcommand{\kms}{\ensuremath{\,\textrm{km s}^{-1}\,{}}}
\newcommand{\msun}{\ensuremath{\,\textrm{M}_{\odot}}}

\newcommand{\mobse}{{\sc mobse}}
\newcommand{\bse}{{\sc bse}}

\definecolor{olivegreen}{RGB}{10,170,10}

\received{XXXX}
\revised{YYYY}
\accepted{ZZZZ}
\submitjournal{ApJ}

\shorttitle{Revising natal kicks}
\shortauthors{Giacobbo \&{} Mapelli}

\begin{document}

\title{Revising natal kick prescriptions in population synthesis simulations}

\correspondingauthor{Nicola Giacobbo}
\email{giacobbo.nicola@gmail.com}

\author{Nicola Giacobbo}
\affiliation{Dipartimento di Fisica e Astronomia ''G. Galilei'', Universit\`a di Padova, vicolo dell'Osservatorio 3, 35122 PD, Italy}
\affiliation{INAF, Osservatorio Astronomico di Padova, vicolo dell'Osservatorio 5, 35122, PD, Italy}
\affiliation{INFN, Sezione di Padova,  via Marzolo 8,  35131, PD, Italy}

\author{Michela Mapelli}
\affiliation{Dipartimento di Fisica e Astronomia ''G. Galilei'', Universit\`a di Padova, vicolo dell'Osservatorio 3, 35122 PD, Italy}
\affiliation{INAF, Osservatorio Astronomico di Padova, vicolo dell'Osservatorio 5, 35122, PD, Italy}
\affiliation{INFN, Sezione di Padova,  via Marzolo 8,  35131, PD, Italy}



\begin{abstract}
  Natal kicks are matter of debate and significantly affect the merger rate density  of compact objects. Here, we present a new simple formalism for natal kicks of neutron stars (NSs) and black holes (BHs). We describe the magnitude of the kick as $v_{\rm kick}\propto{}f_{\rm H05}\,{}\,{}m_{\rm ej}\,{}\,{}m_{\rm rem}^{-1}$, where $f_{\rm H05}$ is a normalization factor, drawn from a Maxwellian distribution with one-dimensional root-mean-square velocity $\sigma{}=265$~km~s$^{-1}$, $m_{\rm ej}$ is the mass of the supernova (SN) ejecta and $m_{\rm rem}$ is the mass of the compact object. This formalism matches the proper motions of young Galactic pulsars and can naturally account for the differences between core-collapse SNe of single stars, electron-capture SNe and ultra-stripped SNe occurring in interacting binaries. Finally, we use our new kick formalism to estimate the local merger rate density of binary NSs ($R_{\rm BNS}$), BH--NS binaries ($R_{\rm BHNS}$) and binary BHs ($R_{\rm BBH}$), based on the cosmic star formation rate density and metallicity evolution. In our fiducial model, we find $R_{\rm BNS}\sim{}600$~Gpc$^{-3}$~yr$^{-1}$, $R_{\rm BHNS}\sim{}10$~Gpc$^{-3}$~yr$^{-1}$ and $R_{\rm BBH}\sim{}50$~Gpc$^{-3}$~yr$^{-1}$, fairly consistent with the numbers inferred from the LIGO-Virgo collaboration.
\end{abstract}

\keywords{binaries: general --- stars: black holes --- stars: neutron --- gravitational waves}


\section{Introduction} \label{sec:Intro}

Compact objects are thought to obtain a spatial velocity at their birth (natal kick), as a result of asymmetric supernova (SN) explosions \citep[e.g][]{1994A&A...290..496J,1996PhRvL..76..352B} or anisotropic emission of neutrinos \citep[e.g.][]{1987IAUS..125..255W,1993A&AT....3..287B,2006ApJS..163..335F,2008PhRvD..77l3009K,2008A&A...489..281S,2014ApJ...792...96T,2019ApJ...880L..28N}. In addition, if the SN occurs in a binary star,  we expect the so-called Blaauw kick to affect the orbital properties of the binary system, even if mass loss is completely symmetric \citep{1961BAN....15..265B}.

Most observational estimates of natal kicks come from pulsar proper motions \citep{1994Natur.369..127L,1997MNRAS.291..569H,2002ApJ...568..289A,2005MNRAS.360..974H,2006ApJ...643..332F}. The kick distribution we can infer from these data is still matter of debate. \cite{2005MNRAS.360..974H} study proper motions of 233 Galactic pulsars. Restricting their analysis to the 73 pulsars younger than $\sim{}3$ Myr (whose proper motions were less affected by the environment), they fit a Maxwellian distribution to the natal kick velocity, with one dimensional root-mean square (rms) velocity $\sigma{}=265$ \kms.

Other works suggest a bimodal velocity distribution of pulsars, with a first peak at low velocities (e.g. $\sim{}0$ \kms~according to \citealt{1998ApJ...496..333F} or $\sim{}90$ \kms~according to \citealt{2002ApJ...568..289A}) and a second peak at high velocities ($\sim{}500$ \kms~according to \citealt{2002ApJ...568..289A}, or even $>600$ \kms, \citealt{1998ApJ...496..333F}). 
Based on VLBI data of 28 isolated pulsars, \cite{2017A&A...608A..57V} also indicate that a double Maxwellian distribution provides a significantly better fit to the observed velocity distribution than a single Maxwellian distribution.

\cite{2016MNRAS.456.4089B} follow a different approach: they focus on binary neutron stars (BNSs) only and find a strong preference for small mass ejection ($\leq{}0.5$ M$_\odot$) and small natal kicks ($v_{\rm kick}\leq{}30$ \kms). Similarly, from the analysis of $r-$process material in ultra-faint dwarf galaxies \cite{2016ApJ...829L..13B} find further support for a prevalence of small natal kicks in BNSs.

The situation for black hole (BH) natal kicks is even more uncertain (see e.g. \citealt{1995MNRAS.277L..35B,1999A&A...352L..87N,2001Natur.413..139M,2002A&A...395..595M,2003Sci...300.1119M,2005ApJ...618..845G,2009ApJ...697.1057F, 2012MNRAS.425.2799R,2017MNRAS.467..298R,2014ApJ...790..119W}). While recent studies (e.g. \citealt{2017MNRAS.467..298R,2017PhRvL.119a1101O,2019arXiv190807199A,2019MNRAS.485.2642G}) suggest that several Galactic BHs received a relatively high natal kick ($\sim{}100$ \kms), we are still far from inferring a distribution of BH kicks from observations.

From a theoretical perspective, hydrodynamical simulations of SN explosions have successfully shown that explosion asymmetries may arise from non-radial hydrodynamic instabilities in the collapsing core \citep{2006AAS...20915006B,2006A&A...457..963S,2007ApJ...654.1006F,2015PASA...32....9F,2012ARNPS..62..407J,2013MNRAS.434.1355J}. Hydrodynamical simulations show that large kick magnitudes can be achieved \citep{2013A&A...552A.126W}, similar to the ones reported by \cite{2005MNRAS.360..974H}. Recently, \cite{2017ApJ...837...84J}, using the gravitational tug-boat mechanism in asymmetric neutrino-driven core-collapse SNe (CCSNe), derived a simple scaling between the natal kick, the energy of the explosion and the amount of asymmetries.

State-of-the-art population-synthesis simulations build on the results of observational constraints and of hydrodynamical models of SN explosion. Most population-synthesis codes (e.g. \bse, \citealt{2000MNRAS.315..543H,2002MNRAS.329..897H}; {\sc seba}, \citealt{1996A&A...309..179P}; {\sc startrack}, \citealt{2008ApJS..174..223B}; \mobse, \citealt{2017MNRAS.472.2422M,2018MNRAS.474.2959G}; {\sc sevn} \citealt{2019MNRAS.485..889S}) implement neutron star (NS) kicks through the Maxwellian distribution derived by \cite{2005MNRAS.360..974H}. Several codes (e.g. {\sc compas}, \citealt{2018MNRAS.481.4009V}; \mobse{}, \citealt{2018MNRAS.480.2011G, 2019MNRAS.482.2234G}, and {\sc combine}, \citealt{2018MNRAS.481.1908K}) assume different Maxwellian distributions, based on the SN mechanism  (usually with a higher rms velocity for CCSNe and a smaller  rms velocity for ECSNe and ultra-stripped SNe).

  The treatment of BH kicks in population-synthesis simulations  depends on whether the BH forms via fallback or via direct collapse. Lighter BHs, that are thought to form via fallback and to receive larger kicks (e.g. \citealt{2013MNRAS.434.1355J}), are assigned natal kicks drawn from the same distribution as NS kicks \citep{2005MNRAS.360..974H}, but corrected either for linear momentum conservation (e.g. \citealt{2013MNRAS.429.2298M,2014MNRAS.441.3703Z})  or for the effect of fallback \citep{2012ApJ...749...91F}. Finally, if massive BHs are allowed to form by direct collapse, no kick is usually assumed apart from the Blaauw mechanism \citep{2012ApJ...749...91F}.

Several recent studies suggest that this approach is not sufficient to capture the complexity of natal kicks. In particular, \cite{2016MNRAS.461.3747B} and \cite{2018MNRAS.480.5657B} propose a new linear relation between the mass of the ejecta (to account for the effect of asymmetries), divided by the mass of the compact object (to conserve linear momentum), and the natal kick.  
Moreover, natal kicks from electron-capture SNe (ECSNe), which are less energetic than CCSNe, are expected to be significantly low \citep{2006ApJ...644.1063D,2015MNRAS.453.1910S,2018ApJ...865...61G,2019MNRAS.482.2234G}. Furthermore, some stars in close binary systems are predicted to have their outer envelope removed and experience ultra-stripped SNe, i.e. SN explosions of naked helium stars that were stripped by their compact companion \citep{2013ApJ...778L..23T,2015MNRAS.451.2123T}. In this case, the natal kick might be lower, because of the low mass of the ejecta \citep{2015MNRAS.454.3073S,2017ApJ...846..170T,2018MNRAS.481.1908K,2018MNRAS.479.3675M}. Finally, recent population-synthesis studies \citep{2018MNRAS.479.4391M,2018MNRAS.480.2011G,2018MNRAS.474.2937C} suggest that very low kicks ($\leq{}50$ \kms) are crucial to match the high local merger rate density of BNSs inferred from LIGO-Virgo data (110--3840 Gpc$^{-3}$ yr$^{-1}$, \citealt{2019PhRvX...9c1040A}).

Here, we propose a new simple prescription for natal kicks which is able to account for both large velocities in young isolated pulsars and small kicks in ultra-stripped SNe, ECSNe and failed SNe. Building upon \cite{2016MNRAS.461.3747B}, we start from the idea that the effect of asymmetries scales with the mass of the ejecta ($m_{\rm ej}$). From linear momentum conservation, we include the dependence of the kick on compact object mass ($m_{\rm rem}$). As a normalization, we take the Maxwellian distribution by \cite{2005MNRAS.360..974H}.

Hence, our new prescription can be written in the form
$v_{\rm kick}\propto{} f_{\rm H05}\,{}m_{\rm ej}\,{}m_{\rm rem}^{-1}$, where $f_{\rm H05}$ is the kick extracted from a Maxwellian with one-dimensional rms $\sigma{}=265$~\kms. 
 For NSs formed from single stars, our formula is basically indistinguishable from \cite{2005MNRAS.360..974H}. For NSs that form in close binaries (going through ECSNe or ultra-stripped SNe), this formalism automatically produces very low kicks, consistent with \cite{2016MNRAS.456.4089B} and \cite{2018MNRAS.479.4391M}. Finally, low-mass BHs (which form through fallback) tend to have significantly larger kicks than massive BHs, formed via direct collapse.

 This paper is organized as follows. In Section~\ref{sec:Code}, we describe our new prescriptions for natal kicks, as implemented in \mobse{}. Then, we show the effect of our new prescriptions on the distribution of natal kicks (Section~\ref{sec:Results}) and we discuss their impact on the merger rate (Section~\ref{sec:Discussion}). Finally, we summarize our results in Section~\ref{sec:Summary}.

\section{Numerical method} \label{sec:Code}
We implement the new prescriptions for natal kicks in our population synthesis code \mobse{}, which is an updated and customized version of \bse~\citep{2000MNRAS.315..543H,2002MNRAS.329..897H}. Here we briefly summarize the main differences between \mobse{} and \bse{} and we refer to previous papers for more details \citep{2018MNRAS.474.2959G,2018MNRAS.480.2011G}.

\subsection{\mobse}
Mass loss by stellar winds of massive hot stars is described in \mobse{} as $\dot{M}\propto{}Z^{\beta}$, where $\beta=0.85, 2.45-2.4\,{}\Gamma_e,$ and $0.05$ for electron-scattering Eddington ratio $\Gamma_e\le{}2/3$, $2/3<\Gamma_e\leq{}1$,  and $\Gamma_e>1$, respectively (see \citealt{2018MNRAS.474.2959G} and references therein).

ECSNe are modeled as described in \cite{2019MNRAS.482.2234G}. In particular, a star with a Helium core mass at the base of the asymptotic giant branch $1.6 \leq{} M_{\rm BABG}/{\rm M}_{\odot} < 2.25$ forms a partially degenerate Carbon-Oxygen core. If this star forms a degenerate Oxygen-Neon core that reaches a mass $M_{\rm ECSN} = 1.38~M_\odot$, it collapses due to the electron-capture on $^{24}$Mg and $^{20}$Ne (see e.g. \cite{2012ApJ...749...91F}).

CCSNe are described as in \cite{2012ApJ...749...91F}, including both the rapid and the delayed model.  The SN shock is launched $<250$ ms and $>500$ ms after the onset of core collapse in the rapid model and in the  delayed model, respectively. This leads to a substantial difference in the energy released by the CCSN. In both models, the mass of the compact object formed via a CCSN is determined by the final mass of the carbon-oxygen core $m_{\rm CO}$: if $m_{\rm CO}\ge{}11$ M$_\odot$ the star collapses to a BH directly (without mass loss), otherwise the  details of the remnant mass depend on the mass of the proto-NS $m_{\rm proto}$ and on the amount of fallback. Specifically, in the rapid model the
mass of the compact object is $m_{\rm rem} = m_{\rm proto} + m_{\rm fb}$,
where $m_{\rm fb} = f_{\rm fb}\,{}(m_{\rm fin} -  m_{\rm proto})$ is the mass accreted by fallback ($m_{\rm fin}$ is the mass of the star at the onset of core collapse and $f_{\rm fb}$ is the fallback parameter, as defined in  \citealt{2012ApJ...749...91F}).

In this work, we introduce a small but crucial difference with respect to the previous versions of \mobse{}: the mass of the proto-NS in the rapid model is $m_{\rm proto}=1.1$ M$_\odot$, while in \cite{2012ApJ...749...91F} and in the previous versions of \mobse{} we adopted $m_{\rm proto}=1.0$ M$_\odot$. This change is fundamental to match the mass of observed NSs \citep{2017ApJ...846..170T}, because with $m_{\rm proto}=1.0$ M$_\odot$ we drastically overestimated the fraction of NSs with mass $<1.2$ M$_\odot$ (see e.g. \citealt{2018MNRAS.480.2011G}).

Finally, \mobse{} includes a treatment for pair instability and pulsational pair instability based on \cite{2017MNRAS.470.4739S}: if the He core mass is $135 \ge{} m_{\rm He} /{\rm M}_\odot \ge{} 64$, the star undergoes a
pair-instability SN and leaves no compact object; if the He core mass
is $64 > m_{\rm He} /{\rm M}_\odot \ge{} 32$, the star undergoes pulsational
pair instability and the final mass of the compact object
is calculated as $m_{\rm rem} = \alpha{}_{\rm P}\,{} m_{{\rm no}\,{}{\rm PPI}}$, where $m_{{\rm no}\,{}{\rm PPI}}$  is the
mass of the compact object we would have obtained if
we had not included pulsational pair instability in our
analysis (just CCSN) and $\alpha{}_{\rm P}$ is a fitting parameter described in \cite{2019arXiv190901371M}.

Other changes with respect to \bse{} include the modeling of core radii (according to \citealt{2014MNRAS.444.3209H}),  the treatment of common envelope (CE, we assume that all Hertzsprung-gap donors merge during CE) and the maximum stellar mass (we extend the mass range up to 150~M$_{\odot}$, \citealt{2016MNRAS.459.3432M}). Apart from the changes summarized in this section, single and binary evolution in \mobse{} is the same as described in \cite{2000MNRAS.315..543H} and \cite{2002MNRAS.329..897H}.

\subsection{Natal kick prescriptions} 

To develop the new kick prescriptions, we start from assuming that the Maxwellian distribution derived by \cite{2005MNRAS.360..974H} is a description of NS kicks from single star evolution. This assumption is a simplification, as we neglect that the \cite{2005MNRAS.360..974H} sample contains not only single pulsars but also pulsar binaries. On the other hand, the 46 young pulsars whose two dimensional proper motions were used by \cite{2005MNRAS.360..974H} to derive  the Maxwellian distribution fit with $\sigma=265$ \kms  are all single pulsars (see Figure 7 in \citealt{2005MNRAS.360..974H}). Another caveat to keep in mind is that a fraction of these single pulsars might have been members of a binary system before their formation. Thus, if new results about proper motions of young single pulsars become available and suggest a significantly different fitting function, we can easily update our prescriptions to include the new fitting function.

Furthermore, we include in our prescriptions the mass of the ejecta $m_{\rm ej}$, because it is reasonable to assume that the magnitude of the kick depends on the total mass ejected during the SN explosion. Finally, to satisfy linear momentum conservation, we also include a term depending on the mass of the compact object $m_{\rm rem}$.

Hence, the new prescription we adopt for SN kicks can be expressed as
\begin{equation}\label{eq:fundamental}
  v_{\rm kick} = f_{\rm H05}\,{} \frac{m_{\rm ej}}{\langle{}m_{\rm ej}\rangle{}}\,{} \frac{\langle{}m_{\rm NS}\rangle{}}{m_{\rm rem}},
\end{equation}
where $f_{\rm H05}$ is a random number extracted from a Maxwellian distribution with one-dimensional rms $\sigma{}=265$ km s$^{-1}$ \citep{2005MNRAS.360..974H}, $\langle{}m_{\rm NS}\rangle{}$ is the average NS mass  and $\langle{}m_{\rm ej}\rangle{}$ is the average mass of the ejecta associated with the formation of a NS of mass $\langle{}m_{\rm NS}\rangle{}$ from single stellar evolution. In our calculations, we adopt $\langle{}m_{\rm NS}\rangle{}=1.2$ \msun{} and $\langle{}m_{\rm ej}\rangle{}=9$ \msun{}, respectively. These values are calibrated at $Z=0.02$, which is approximately the metallicity of the Milky Way. We compute $m_{\rm ej}$ as the difference between the final mass of the star (before the SN explosion) and the mass of the remnant (including mass loss due to neutrinos). The basic idea behind this normalization is that we want neutron stars formed from single star evolution at solar metallicity to receive a kick consistent with the proper motions of young single pulsars in the Milky Way (which we approximate as $f_{\rm H05}$) . With this normalization, more massive compact objects receive smaller kicks because of linear momentum conservation. Similarly, compact objects that form from binary evolution (where $m_{\rm ej}$ is generally smaller than the average single NS case) also receive a smaller kick than NSs formed from single stars.

To check the impact of compact-object mass on the final kicks, we also run some tests with a second prescription, independent of $m_{\rm rem}$:
\begin{equation}\label{eq:alternative}
  v_{\rm kick} = f_{\rm H05}\,{} \frac{m_{\rm ej}}{\langle{}m_{\rm ej}\rangle{}}.
\end{equation}

\begin{deluxetable}{cc}
\tablenum{1}
\tablecaption{Models. \label{tab:sims}}
\tablewidth{0pt}
\tablehead{
	\colhead{\bf ID} & \colhead{\bf Natal kicks}} 
\startdata
			Ej1 & \begin{tabular}{@{}c@{}} $\sigma$ = 265 km s$^{-1}$, eq.~\ref{eq:fundamental} \end{tabular} \\ 
			Ej2 & \begin{tabular}{@{}c@{}} $\sigma$ = 265 km s$^{-1}$, eq.~\ref{eq:alternative} \end{tabular} \\
			H05 & \begin{tabular}{@{}c@{}} $\sigma$ = 265 km s$^{-1}$, eq~\ref{eq:fryer} \end{tabular}\\
			$\sigma15$ & \begin{tabular}{@{}c@{}} $\sigma$ = 15 km s$^{-1}$, eq~\ref{eq:fryer} \end{tabular} \\
                        \enddata
	\tablecomments{Column 1: name of the simulation; column 2: Natal-kick prescription.}
	
\end{deluxetable}

These prescriptions have several advantages. Firstly, they are simple to implement in population-synthesis codes. Secondly, they are quite universal: they can be used for both NSs and BHs, for both single and binary star evolution, for both ECSNe and CCSNe (including the case of ultra-stripped SNe).


\begin{figure*}	 	
\fig{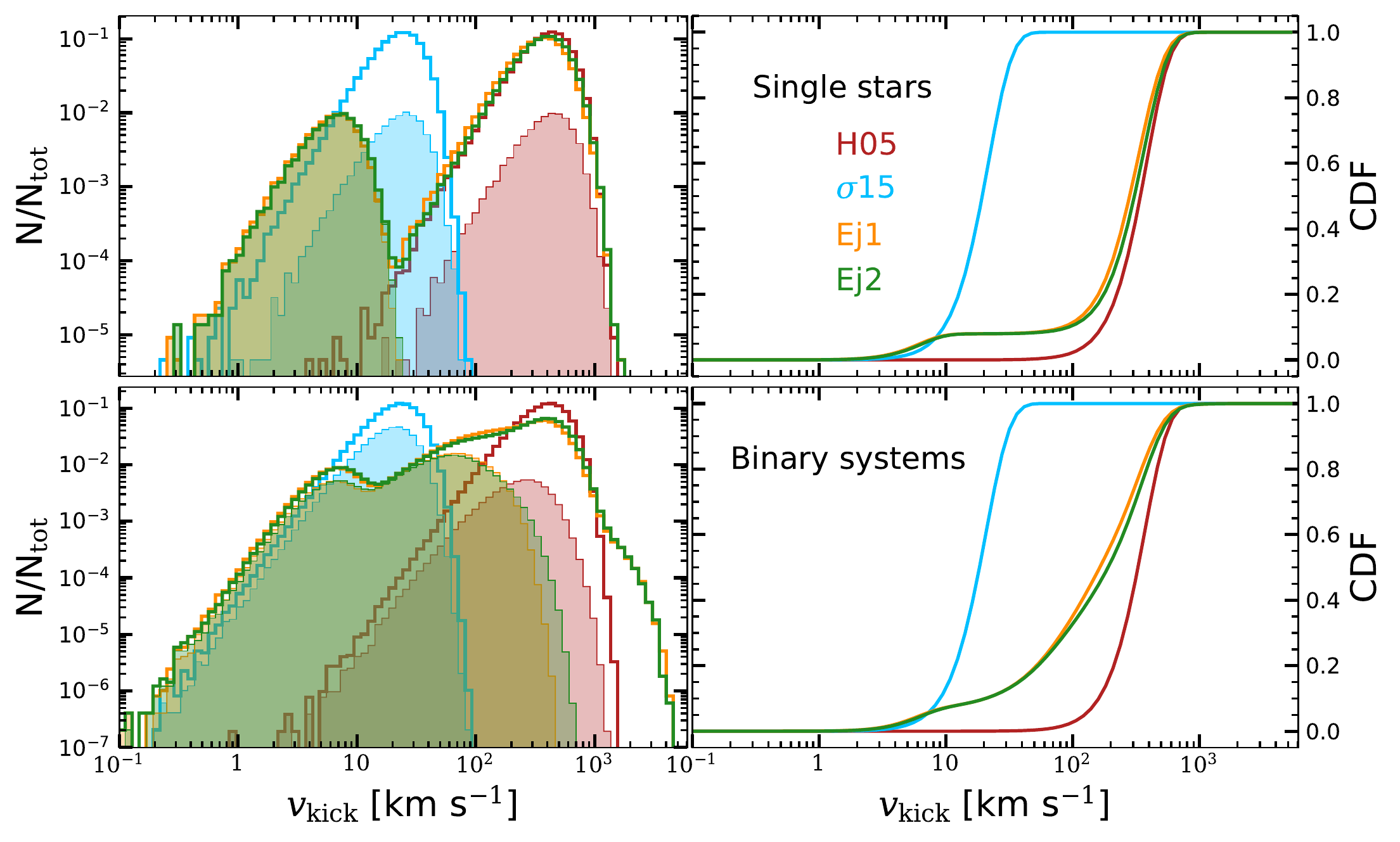}{0.8\textwidth}{} 
\caption{Left-hand panels: distribution of natal kicks for all NSs formed from single stars (top) and for those formed from binary systems (bottom) at $Z=0.02$. Orange line: model Ej1; green: Ej2; red: H05; blue: $\sigma{}15$. The filled histograms represent the subset of NSs formed via ECSNe (top) and the subset of NSs that are still gravitationally bound to their companion after the SN (bottom). Right-hand panels: cumulative distribution function (CDF) of  natal kicks for all NSs. \label{fig:NSdistributions}}
\end{figure*}

\begin{figure*}
\fig{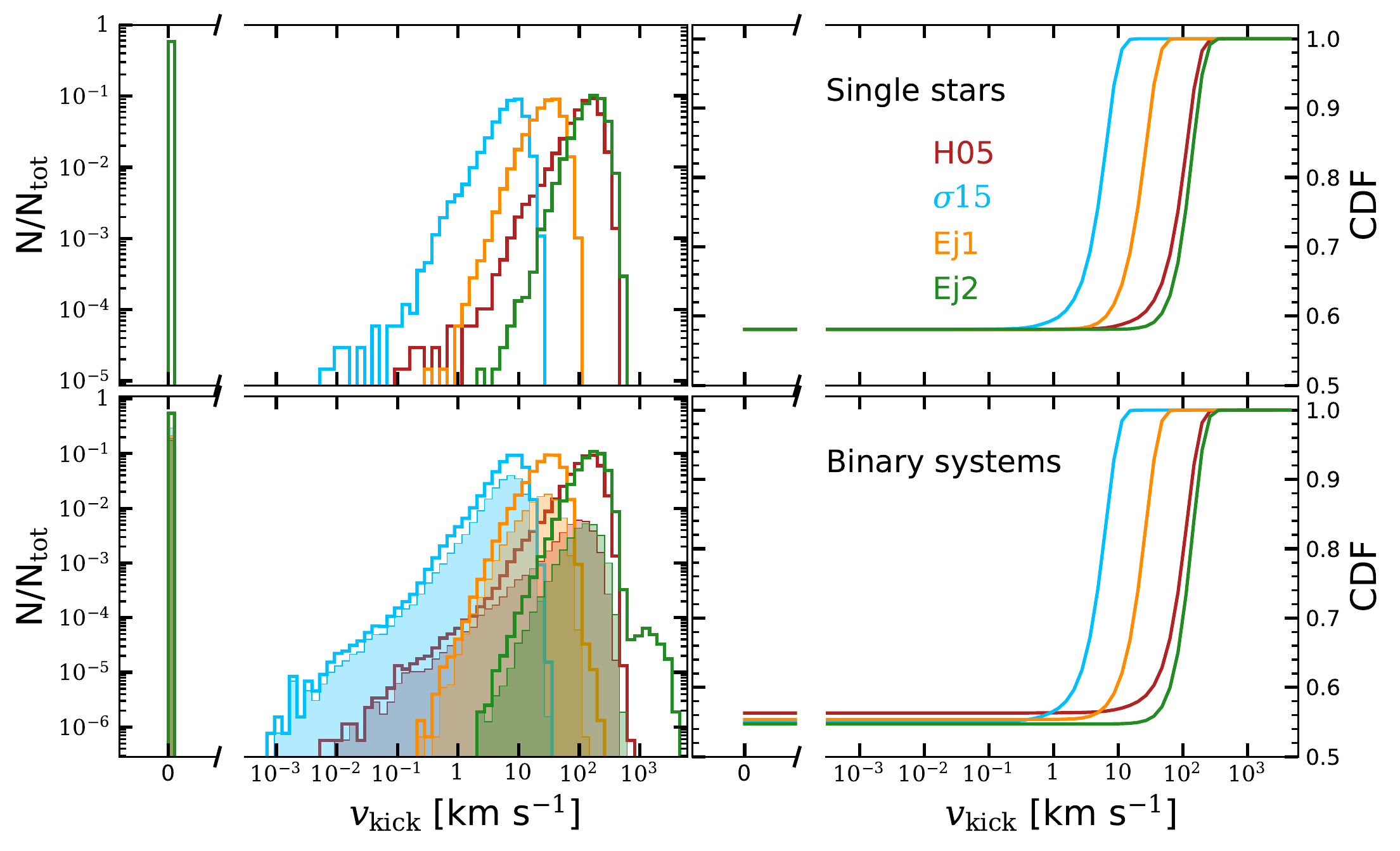}{0.8\textwidth}{} 
\caption{Same as Figure~\ref{fig:NSdistributions}, but for BHs formed from single star evolution (top) and from binary star evolution (bottom)  at $Z=0.02$. The break on the x-axis allows to show BHs with zero natal kick (formed from direct collapse).\label{fig:BHdistributions}}
\end{figure*}

\subsection{Simulation setup}
We used \mobse~to simulate a large set of both single stars and binary systems. For single stars, and for the primary star in binary systems, we randomly draw the initial mass ($m_1$) from a Kroupa initial mass function \citep{2001MNRAS.322..231K} 
$\mathfrak{F}(m_1) ~\propto~ m_1^{-2.3}$ with $m_1 \in [5-150]\msun$. 
The mass of the stellar companion in binaries is derived from the mass ratio as $\mathfrak{F}(q)~ \propto ~q^{-0.1}$ with $q = m_2/m_1 \in [0.1-1]$ (following \citealt{2012Sci...337..444S}). Finally, the eccentricity $e$ and the orbital period $P$ are also drawn from the distributions proposed by \citet{2012Sci...337..444S}:  
	$\mathfrak{F}(e) ~\propto ~e^{-0.42}$ (with $0\leq e < 1$) and 
	$\mathfrak{F}(P) \propto (P)^{-0.55}$ (with $P = \log_{10}(P/{\rm day}) \in [0.15-5.5]$).

We assume the rapid model for CCSNe \citep{2012ApJ...749...91F}. We assume CE efficiency $\alpha{}=5$ (unless otherwise stated) and we derive $\lambda{}$ from the formulas in \cite{2014A&A...563A..83C}. In appendix~\ref{sec:app}, we discuss the impact of different choices of $\alpha{}$ on our main results.

\vspace{0.5cm}

We have run the following four sets of simulations (see Table~\ref{tab:sims}). 
\begin{itemize}
\item[Ej1:] natal kicks are implemented  as in equation~\ref{eq:fundamental};
\item[Ej2:] natal kicks are drawn from equation~\ref{eq:alternative};
\item[H05:] natal kicks are generated from a Maxwellian with $\sigma{}=265$ \kms~for both CCSNe and ECSNe (see model EC265$\alpha$5 in \citealt{2019MNRAS.482.2234G}), plus a correction for the amount of fallback following \citealt{2012ApJ...749...91F} (see below equation~\ref{eq:fryer});

\item[$\sigma$15:] natal kicks are drawn from a single Maxwellian with rms$=15$~\kms~for both ECSNe and CCSNe (see model CC15$\alpha$5 in \citealt{2018MNRAS.480.2011G}), plus a correction for the amount of fallback as in \cite{2012ApJ...749...91F}.
\end{itemize}

The correction for the amount of fallback in models H05 and $\sigma{}15$ is implemented as follows. We draw the natal kick as
\begin{equation}\label{eq:fryer}
  v_{\rm kick}= (1-f_{\rm fb})\,{}f_{\rm H05},
\end{equation}
  where $f_{\rm H05}$ is a random number drawn from the Maxwellian distribution, while $f_{\rm fb}$ is the fallback fraction, defined as  $f_{\rm fb}=m_{\rm fb}/(m_{\rm fin}-m_{\rm proto})$, where $m_{\rm fin}$ is the mass of the star at the onset of core collapse and  $m_{\rm fb}$ is the mass that falls back and is accreted by the proto-NS \citep{2012ApJ...749...91F}. The main difference between our new prescriptions and equation~\ref{eq:fryer} is that the latter does not depend significantly on the mass of the ejecta (in equation~\ref{eq:fryer} $v_{\rm kick}\propto{}m_{\rm ej}/m_{\rm fin}$, i.e. $m_{\rm fin}$ compensates the impact of $m_{\rm ej}$). 

The direction of the kicks has been computed adopting the same prescriptions as implemented in {\sc{BSE}} and described in the appendix of \cite{2002MNRAS.329..897H}. In particular, kicks are assumed to be isotropically oriented over the sphere. After randomly drawing the kick direction under this assumption, we calculate how the kick affects the orbital elements and check whether the system remains bound after the supernova explosion.

For each set of simulations we consider 12 different metallicities: $Z =0.0002$, $0.0004$, $0.0008$, $0.0012$, $0.0016$, $0.002$, $0.004$, $0.006$, $0.008$, $0.012$, $0.016$ and $0.02$. For each metallicity, we simulated $10^7$ binary systems and $5 \times 10^5$ single stars. Thus, for each model we simulate $1.2\times 10^8$ massive binaries and $6 \times 10^6$ single stars.

\begin{deluxetable}{cccc}
\tablenum{2}
\tablecaption{Median values of natal kicks. \label{tab:table2}}
\tablewidth{0pt}
\tablehead{
  \colhead{Model}  & \colhead{NS/BH} 	& \colhead{Progenitor star} & \colhead{$\tilde{v}_{\rm kick}$~(\kms{})}
}
\startdata
Ej1 & NS & single & 322\\
Ej1 & NS & binary & 188\\
Ej2 & NS & single & 351\\
Ej2 & NS & binary & 218\\
H05 & NS & single & 392\\
H05 & NS & binary & 375\\
$\sigma{}15$ & NS & single & 22\\
$\sigma{}15$ & NS & binary & 21\vspace{0.3cm}\\
Ej1 & BH & single & 30\\
Ej1 & BH & binary & 30\\
Ej2 & BH & single & 164\\
Ej2 & BH & binary & 165\\
H05 & BH & single & 127\\
H05 & BH & binary & 129\\
$\sigma{}15$ & BH & single & 7\\
$\sigma{}15$ & BH & binary & 7\\
\enddata
\tablecomments{Column 1: model; column 2: compact-object type (NS or BH); column 3: whether the progenitor star was born as a single or a binary star; column 4: median value of natal kicks.}	
\end{deluxetable}

\section{Results}\label{sec:Results}
\subsection{Natal kicks in single stars}
The top panels of Figure~\ref{fig:NSdistributions} show the natal kick distribution of NSs born from single stars with solar metallicity ($Z=0.02$). 
NS kicks from simulations Ej1 and Ej2 are extremely similar to each other. They both show two different peaks, one centered at $\sim{}400-450$~\kms{} and produced by CCSNe, the other centered at $\sim{}6-8$~\kms{} and produced by ECSNe. This happens because $m_{\rm ej}$ of ECSNe is significantly smaller than that of CCSNe, leading to smaller kicks. Thus, our new prescriptions are able to distinguish between CCSN kicks and ECSN kicks, without the need for a separate treatment. 

The distribution of NS kicks from CCSNe in simulation H05 (drawn from a single Maxwellian with $\sigma{}=265$~km~s$^{-1}$) is remarkably similar to the peak produced by CCSNe in simulations Ej1 and Ej2. This confirms that simulations Ej1 and Ej2 are a good match to the fit by \cite{2005MNRAS.360..974H} for large NS kicks. On the other hand, runs Ej1 and Ej2 can also naturally reproduce the low kicks of ECSNe. 
Finally, simulation $\sigma{}15$ produces single NS kicks that are significantly lower than the other runs, unable to explain a large fraction of the sample by \cite{2005MNRAS.360..974H}.

The top panels of Figure~\ref{fig:BHdistributions} show the natal kick distribution of BHs born from single stars with solar metallicity ($Z=0.02$). All the four models predict that $\sim{}60$~\% of BHs receive approximately no kick, because their progenitors collapse to a BH directly, without SN explosions. The remaining BHs receive a kick. Models H05 and Ej2 predict the largest maximum kicks, up to $\sim{}450$ and $\sim{}550$ \kms{}, respectively. 
In fact, the kick prescriptions in H05 and Ej2 do not depend on compact-object mass. Model $\sigma{}15$ predicts the lowest BH kicks (up to $\sim{}30$ km s$^{-1}$), while model Ej1 ($v_{\rm kick}\leq{}100$~km s$^{-1}$) is intermediate between the considered models, thanks to the dependence on $m_{\rm rem}$.

\subsection{Natal kicks in binary stars}
The bottom panels of Figure~\ref{fig:NSdistributions} (Figure~\ref{fig:BHdistributions}) shows the natal kicks of NSs  (BHs) formed from the evolution of binary stars with $Z=0.02$. Binary evolution significantly affects the distribution of NS natal kicks in all models and especially in run Ej1 and Ej2. The Kolmogorov-Smirnov (KS) test confirms that the probability that natal kicks of NSs formed from single stars and from binary evolution are drawn from the same distribution is nearly zero ($<10^{-20}$).  Table~\ref{tab:table2} shows that the median value of NS kicks is significantly lower for binary stars than for single stars  in models Ej1 and Ej2. In general, binary evolution tends to increase the number of NSs with small kicks, because dissipative mass transfer tends to reduce $m_{\rm ej}$.  
 On the other hand, binary evolution also triggers the formation of a few NSs with even larger kicks than in the case of single star evolution. 

Binary evolution has a smaller impact on NS kicks in models H05 and $\sigma{}15$ by construction (see Table~\ref{tab:table2}), because they do not depend significantly on $m_{\rm ej}$. The only effect of binary evolution on models H05 and $\sigma{}15$ is that mass transfer can change $m_{\rm rem}$ and the amount of fallback, hence affecting natal kicks. This affects mostly BHs, while it has negligible impact on NSs. 

The distribution of NS kicks in simulations Ej1 and Ej2 are very similar to each other, even when we account for binary evolution. As expected, NSs that remain members of a binary system after the kick (filled histograms) have significantly smaller kicks than single NSs in runs Ej1, Ej2 and H05. In model Ej1 (Ej2), the maximum kick undergone by NSs that remain in binaries is $v_{\rm kick}\sim{}400$~\kms ($\sim{}600$~\kms{}), while the maximum possible NS kick is $v_{\rm kick}\sim{}4500$~\kms ($\sim{}4500$~\kms). The maximum possible NS kick with our new models ($v_{\rm kick}\sim{}4500$~\kms) is extremely unlikely, as less than $\sim{}10^{-4}$~\% of all simulated NSs in binary systems have $v_{\rm kick}\ge{}4000$~\kms (and less than $\sim{}0.03$~\% have $v_{\rm kick}\ge{}2000$~\kms).

Finally, binary evolution has  a different effect on BH kicks. In the case of BHs, dissipative mass transfer affects $m_{\rm rem}$, producing smaller BHs. This explains why the percentage of BHs that undergo no kick decreases (to about five per cent) in all models. Table~\ref{tab:table2} shows that the median value of BH kicks is not affected by the binarity of progenitors.

\begin{figure}
  \fig{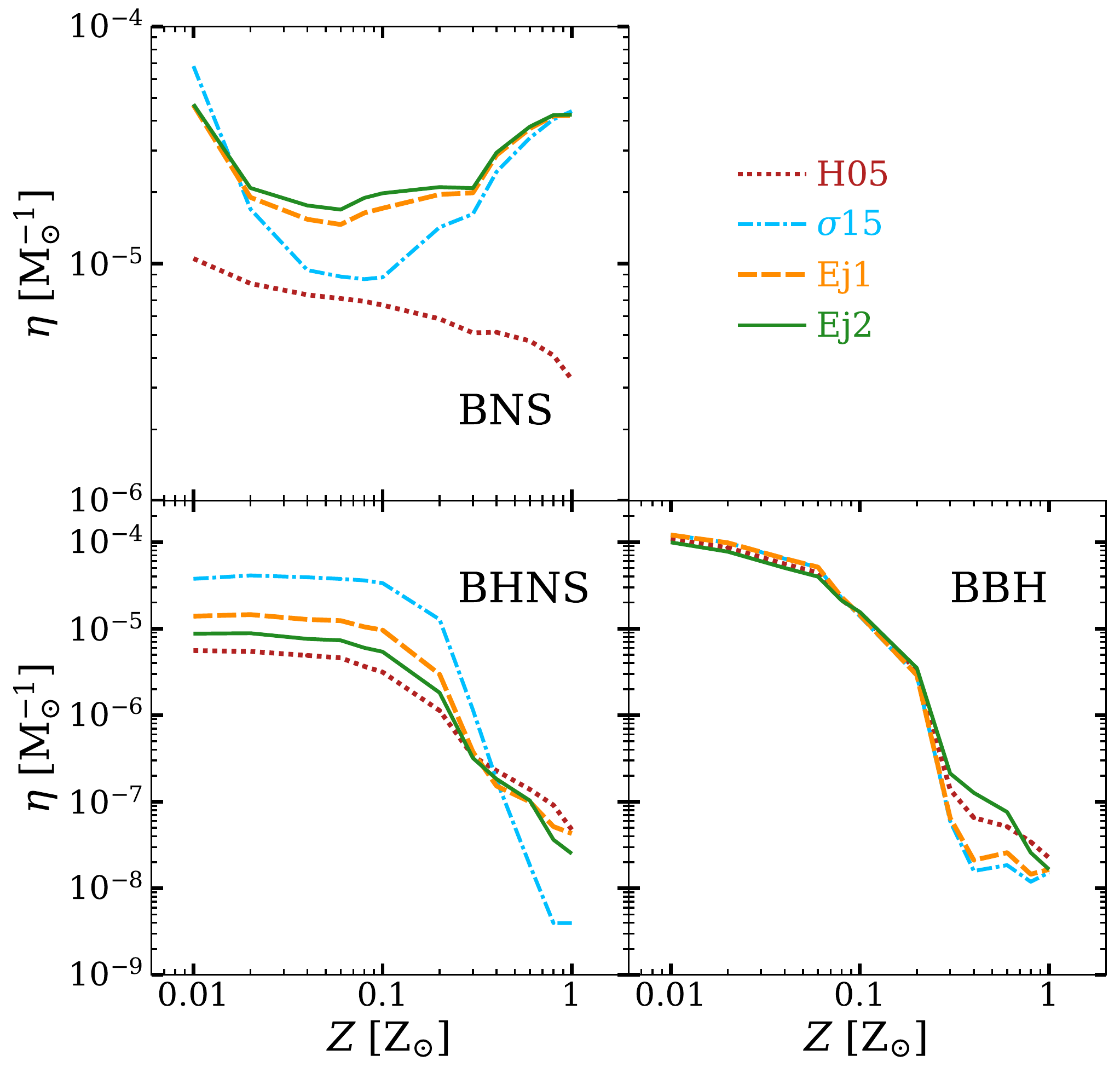}{0.47\textwidth}{}
  \caption{Merger efficiency ($\eta$ from eq.~\ref{eq:eta}) as a function of the progenitor's metallicity for all sets of simulations (see Table~\ref{tab:sims}). {\it Top-left:} BNSs; {\it bottom-left:} BHNSs; {\it bottom-right:} BBHs. \label{fig:eta}}
\end{figure}

\begin{figure}
  \fig{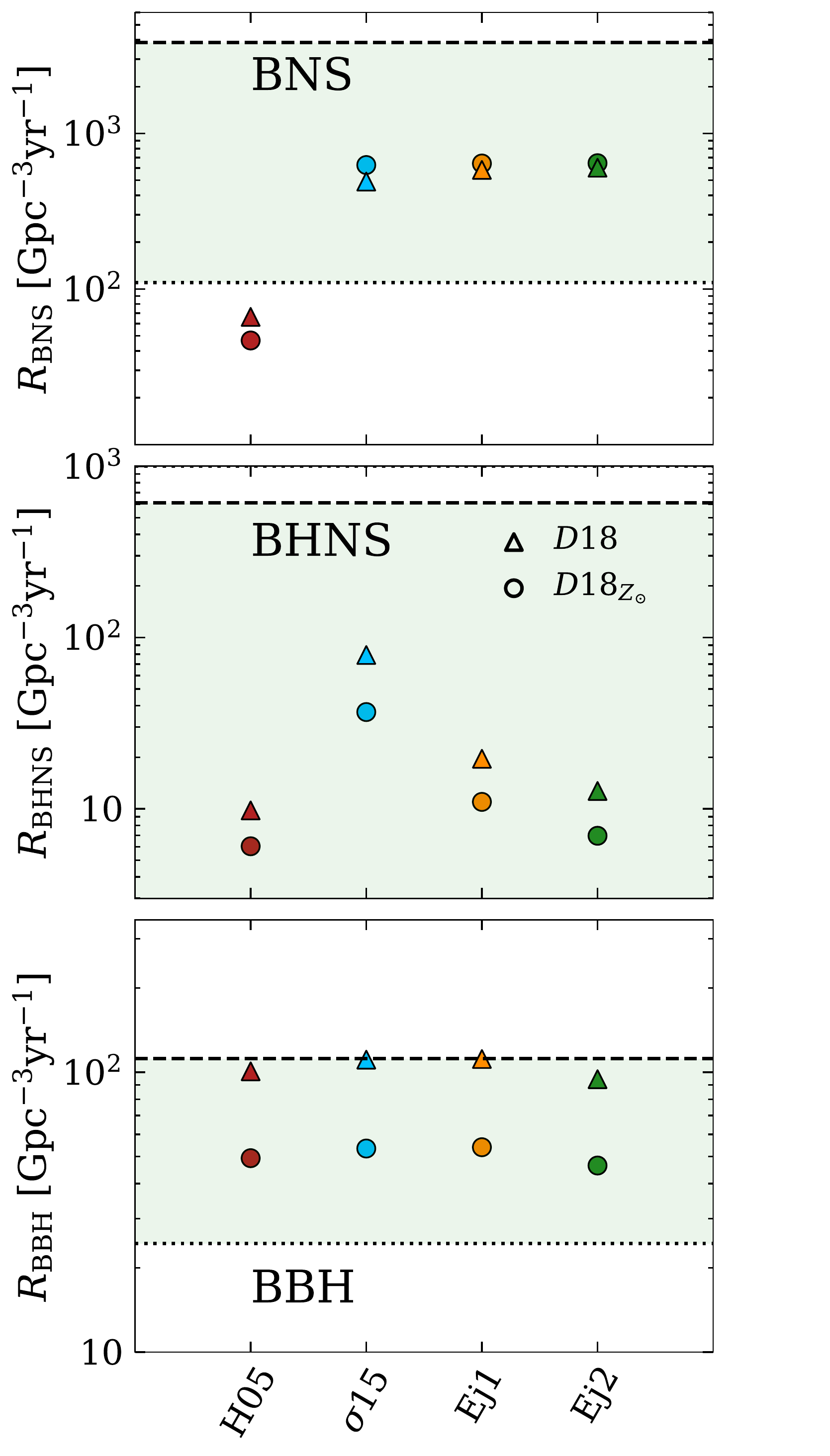}{0.45\textwidth}{}
  \caption{Local merger rate density $R$ from eq.~\ref{eq:rate}.  Top, middle and bottom panel: local merger rate density of BNSs ($R_{\rm BNS}$), BHNSs ($R_{\rm BHNS}$) and BBHs ($R_{\rm BBH}$), respectively. Triangles (circles) assume model $D18$ ($D18_{Z_{\odot}}$) for the cosmic evolution of metallicity. The green shaded regions represent the 90 per cent confident interval of the merger rate density inferred by \cite{2019ApJ...882L..24A} for BBHs and \cite{2019PhRvX...9c1040A} for BNSs and BHNSs.\label{fig:Rloc}}
\end{figure}
\subsection{Merger efficiency}
For each set of binary simulations we compute the merger efficiency, that is the number of compact-object mergers occurring in a given stellar population, integrated over the Hubble time, divided by the total initial stellar mass. As already described in \cite{2017MNRAS.472.2422M}, 
the merger efficiency $\eta{}$ is given by
\begin{equation}\label{eq:eta}
  \eta{} =  f_{\rm bin}\,{} f_{\rm IMF} \,{} \frac{N_{\rm merg}}{M_{\rm tot,sim}},
\end{equation}
where $N_{\rm merg}$ is the number of mergers of binary BHs (BBHs), or BH -- NS binaries (BHNSs), or binary NSs (BNSs), and $M_{\rm tot,sim}$ is the initial total mass of the simulated binary population.  Since we simulated only massive binaries, we introduce two corrections factors: $f_{\rm bin}=0.5$ (to correct for the fact that $\sim{}50$ per cent of stars are single, \citealt{2013A&A...550A.107S}) and $f_{\rm IMF}=0.285$ (to account for the total mass of stars below the minimum mass we simulate).

Figure~\ref{fig:eta} shows $\eta$ as a function of metallicity for all runs (see Table~\ref{tab:sims}). 
The merger efficiency of both BBHs and BHNSs strongly depends on metallicity: BH mergers are at least two orders of magnitude more common in a metal-poor population than in a metal-rich one. This result is well known and is consistent with previous work \citep{2013ApJ...779...72D,2018A&A...619A..77K,2018MNRAS.474.2959G,2018MNRAS.480.2011G}.  The merger efficiency of BNSs depends only mildly on metallicity. 
The decrease of $\eta{}$ at intermediate metallicity ($0.0004\lesssim Z\lesssim 0.04$) in the models with relatively low kicks (Ej1, Ej2 and $\sigma{}15$) is caused by premature mergers of the progenitor stars, because stellar radii during the Hertzsprung gap and the red giant phase are larger at intermediate metallicity (see \citealt{2018MNRAS.480.2011G,2019MNRAS.485..889S}). In model~H05, $\eta{}$ decreases with increasing metallicity, because  the ability of CE to shrink the binary becomes decisive when SN kicks are high: at high metallicity stars lose their envelope quite effectively, reducing the impact of CE.

More importantly, Figure~\ref{fig:eta} shows that our new kick prescriptions (models Ej1, Ej2) produce approximately the same BNS merger efficiency as model $\sigma{}15$, which assumes unrealistically small kicks.  
For BHNSs, the new kick prescriptions give a merger rate efficiency more similar to H05 than to $\sigma{}15$. Finally, the merger efficiency of BBHs is not significantly affected by the new kick prescriptions, because most merging BBHs receive no kick (or very small kick) in all considered models.

\subsection{Local merger rate}
Following \cite{2018MNRAS.480.2011G} and \cite{2019MNRAS.485..889S}, we compute the local merger rate density $R$  as 
\begin{equation}\label{eq:rate}
R = \frac{1}{H_0\,t_{\rm lb}(z=0.1)} \int_{z_{\rm max}}^{z_{\rm min}} \frac{f_{\rm loc}(z,\,{}Z)~{\rm SFR}(z)}{(1+z)\,{}\mathcal{E}(z)}dz,
\end{equation}
where ${\rm SFR}(z)$ is the star formation rate density (for which we adopt the fitting formula proposed by \citealt{2017ApJ...840...39M}), $\mathcal{E}(z)=\left[\Omega_{\rm M}\,{}(1+z)^3+\Omega_\Lambda{}\right]^{1/2}$, $t_{\rm lb}(z=0.1)$ is the look-back time at redshift $z=0.1$, and $f_{\rm loc}(z,\,{}Z)$ is the fraction of merging systems that formed at a given redshift $z$ and merge in the local Universe ($z\le{}0.1$) per unit solar mass. We assume $z_{\rm max}=15$ and $z_{\rm min}=0$. Finally, $H_0$, $\Omega_{\rm M}$ and $\Omega_\Lambda{}$ are the cosmological parameters for which we take the values from \cite{2016A&A...594A..13P}.

The term  $f_{\rm loc}(z,\,{}Z)$ clearly depends not only on redshift but also on metallicity (which is important especially for BBHs and BHNSs, see Fig.~\ref{fig:eta}). We derive $f_{\rm loc}(z,\,{}Z)$ directly from the merger efficiency $\eta{}$ (equation~\ref{eq:eta}), by assuming that all stars formed at a given redshift have the same metallicity. We describe the evolution of  metallicity across cosmic time with two different models. In model {\it D18}, the metallicity evolves with redshift as $\log{Z(z)/Z_{\odot{}}}=-0.24\,{}z-0.18$. This formula is the fit to the metallicity evolution of a large sample of damped Lyman-$\alpha$ absorbers (with redshift between 0 and 5) presented in \cite{2018A&A...611A..76D} (see their Figure~4 and Table~1). With respect to previous work (e.g. \citealt{2012ApJ...755...89R}, whose results we used in \citealt{2018MNRAS.480.2011G}), \cite{2018A&A...611A..76D} consider a larger sample of damped Lyman-$\alpha$ absorbers and make a new correction for dust. This allows them to recover a present-day average metallicity  $Z(z=0)\sim{}0.66$ Z$_\odot$ (where we assume $Z_\odot=0.02$), much closer to the solar metallicity than previous work. 

In the second model we adopt ({\it D18$_{Z_{\odot}}$}), the metallicity evolves with redshift as $\log{Z(z)/Z_{\odot{}}}=-0.24\,{}z$. This model is obtained by re-scaling model {\it D18} to obtain $Z(z=0) = Z_{\odot}$. The reason for this re-scaling is that metallicity measurements from galaxies in the Sloan Digital Sky Survey indicate that the average local metallicity is $Z(z=0) \sim{} Z_{\odot}$ \citep{2008MNRAS.383.1439G}.

Figure~\ref{fig:Rloc} shows the local merger rate $R_{\rm BNS}$, $R_{\rm BHNS}$ and $R_{\rm BBH}$ for BNSs, BHNSs and BBHs, respectively, considering both models of  metallicity evolution, namely {\it D18} and {\it D18$_{Z_{\odot}}$}. 

The new kick prescriptions Ej1 and Ej2 produce a BNS merger rate $R_{\rm BNS}\sim{}600$ Gpc$^{-3}$ yr$^{-1}$, consistent with the local merger rate inferred from GW170817 ($R_{\rm GW170817}=110-3840$ Gpc$^{-3}$ yr$^{-1}$, \citealt{2017PhRvL.119p1101A,2019PhRvX...9c1040A}). The rate from Ej1 and Ej2 is very similar to the rate we obtain with the low-kick model $\sigma{}15$ and about one order of magnitude higher than the rate we obtain with model H05. 

All our models are consistent with the upper limit on BHNSs by \cite{2019PhRvX...9c1040A}. Models Ej1 and Ej2 produce rates that are significantly smaller than $\sigma{}15$ and slightly higher than H05. Finally, the merger rate density of BBHs is extremely sensitive to metallicity.
Model {\it D18} results in a factor of $\sim{}2$ higher BBH merger rate than  model {\it D18$_{Z_{\odot}}$}, but still within the 90 per cent credible interval inferred by the LIGO-Virgo collaboration (LVC, $R_{\rm BBH}\sim{}24 - 112$ Gpc$^{-3}$ yr$^{-1}$, \citealt{2019ApJ...882L..24A}). The four kick prescriptions produce approximately the same BBH merger rate density, because all of them suppress natal kicks in massive BHs by approximately the same amount.

\section{Discussion} \label{sec:Discussion}
Recent studies \citep{2018MNRAS.480.2011G,2019MNRAS.482.2234G,2018MNRAS.479.4391M,2018MNRAS.474.2937C,2019MNRAS.482.5012C,2018A&A...615A..91B,2019MNRAS.482.5012C} have shown that it is quite difficult to match the BNS merger rate inferred from GW170817 ($R_{\rm GW170817}$) with state-of-the-art population-synthesis models. 
Models describing natal kicks as in \cite{2005MNRAS.360..974H} produce a merger rate density lower than the range inferred from GW170817. In order to match $R_{\rm GW170817}$, \cite{2018MNRAS.480.2011G} had to introduce model $\sigma{}15$ with very low natal kicks. On the other hand, model $\sigma{}15$ does not match the observed proper motions of young single pulsars \citep{2005MNRAS.360..974H,2017A&A...608A..57V}.

Our new kick prescriptions (models Ej1 and Ej2) solve this tension with data, because they match $R_{\rm GW170817}$ and at the same time they reproduce the natal kicks of young pulsars. Moreover, Ej1 and Ej2 naturally account for the difference between kicks produced by CCSNe of single stars, ECSNe and ultra-stripped SNe in binary stars \citep{2017ApJ...846..170T}. 

The only parameter we need to set to a rather unusual value in order to match $R_{\rm GW170817}$ is the $\alpha{}$ parameter of CE. Our models Ej1 and Ej2 require $\alpha\ge{}3$ to match $R_{\rm GW170817}$ (see Appendix~\ref{sec:app}) and we assume $\alpha{}=5$ as a fiducial value. According to the $\alpha{}-$formalism \citep{1984ApJ...277..355W,1985ibs..book...39W,1990ApJ...358..189D}, values of $\alpha{}>1$ require that additional sources of energy  assist the orbital energy of the system in ejecting the envelope (see \citealt{2013A&ARv..21...59I} for a review). Recently, \cite{2019arXiv190712573F} have presented one-dimensional hydrodynamic simulations of a neutron-star binary  evolving through CE. Their results support very large values of $\alpha{}\approx{}5$, consistent with our work. Once more, this highlights the need for a better physical model of the CE process. Another possibility is that GW170817 was a very lucky event, leading to an overestimate of the local merger rate. A more accurate estimate of the observed merger rate will be available in the next few months, because the third observing run of LVC is currently ongoing.

The key ingredient in our prescriptions is the dependence of $v_{\rm kick}$ on the mass of the ejecta ($v_{\rm kick}\propto{}m_{\rm ej}$). Models adopting the fallback formalism \citep{2012ApJ...749...91F} predict significantly larger kicks for NSs even if they come from ultra-stripped SNe, because in this formalism $v_{\rm kick}\propto{}m_{\rm ej}/m_{\rm fin}$ (i.e. the contribution of $m_{\rm ej}$ to the kick is compensated by the stellar mass $m_{\rm fin}$ at the onset of the SN). The only models that predict a similar behavior to our prescriptions are those presented in \cite{2016MNRAS.461.3747B,2018MNRAS.480.5657B}. \cite{2018MNRAS.480.5657B} derive a BNS merger rate density $R_{\rm BNS}\sim{}3860$ Gpc$^{-3}$ yr$^{-1}$. The difference with respect to our results might arise from the calculation of the local merger rate (\citealt{2018MNRAS.480.5657B} consider only the local SFR, without taking into account the evolution of metallicity across cosmic time) and from different population-synthesis codes.

%
\section{Summary} \label{sec:Summary}
We have proposed a new simple formalism to implement NS and BH kicks in population-synthesis simulations. We describe kick velocities as $v_{\rm kick}\propto{}f_{\rm H05}\,{}m_{\rm ej}\,{}m_{\rm rem}^{-1}$, where $f_{\rm H05}$ is a random number drawn from a Maxwellian distribution with one-dimensional rms $\sigma{}=265$ km s$^{-1}$ \citep{2005MNRAS.360..974H}, $m_{\rm ej}$ is the mass of the ejecta and $m_{\rm rem}$ the mass of the compact object. We have included this formalism in our population-synthesis code \mobse{}.

This formalism can naturally account for the differences between core-collapse SNe (CCSNe) of single stars and electron-capture SNe (ECSNe) or ultra-stripped SNe occurring in binary systems. In fact, CCSNe of single stars have larger values of $m_{\rm ej}$ than ECSNe, ultra-stripped SNe and other SNe occurring in interacting binaries. Hence, the kicks of NSs in interacting binary systems are significantly lower than the kicks of single NSs (Fig.~\ref{fig:NSdistributions} and Table~\ref{tab:table2}).

The kicks of BHs are generally lower than the kicks of NSs (Fig.~\ref{fig:BHdistributions} and Table~\ref{tab:table2}), because $m_{\rm rem}$ is significantly larger and $m_{\rm ej}$ is generally lower than for NSs (in the case of direct collapse $m_{\rm ej}=0$, thus the kick is zero).

We estimate the local merger rate density of BNSs ($R_{\rm BNS}$), BHNSs ($R_{\rm BHNS}$) and BBHs ($R_{\rm BBH}$) with the new kick formalism. The merger rate density of BBHs and BHNSs is extremely sensitive to metallicity evolution. With the new kick prescription Ej1 (Ej2), we find $R_{\rm BBH}\sim{}53$ Gpc$^{-3}$ yr$^{-1}$ ($\sim{}46$ Gpc$^{-3}$ yr$^{-1}$) and $R_{\rm BHNS}\sim{}10$ Gpc$^{-3}$ yr$^{-1}$ ($\sim{}7$ Gpc$^{-3}$ yr$^{-1}$), when adopting model {\it D18$_{Z_\odot}$} for the cosmic evolution of metallicity. These results are consistent with estimates from the LVC \citep{2019PhRvX...9c1040A,2019ApJ...882L..24A}.

The BNS merger rate density depends very mildly on metallicity evolution. With the new kick formalism we estimate $R_{\rm BNS}\sim{}640$ Gpc$^{-3}$ yr$^{-1}$ (adopting model {\it D18$_{Z_\odot}$} for the cosmic evolution of metallicity), consistent with the rate inferred from GW170817 \citep{2019PhRvX...9c1040A}. Interestingly, the BNS merger rate density we find with the new kick prescriptions is extremely close to the one we derived with our previous model $\sigma{}15$ \citep{2018MNRAS.480.2011G}, that assumes extremely low NS kicks (drawn from a Maxwellian with one-dimensional rms $\sigma{}=15$ km s$^{-1}$). Model $\sigma{}15$ matches $R_{\rm GW170817}$ but is in tension with the proper motions of several young Galactic pulsars, while the new kick formalism overcomes this issue.

In conclusion, our new kick formalism is consistent with both observations of proper motions from young Galactic pulsars \citep{2005MNRAS.360..974H} and with the merger rate density of BBHs, BHNSs and BNSs inferred from the LVC \citep{2019PhRvX...9c1040A,2019ApJ...882L..24A}. These results, together with its intrinsic simplicity, make our new kick formalism an interesting prescription for population synthesis simulations.

\acknowledgments

We thank the anonymous referee for their useful suggestions. We also thank Mario Spera, Alessandro Ballone, Mario Pasquato and Filippo Santoliquido for fruitful discussions. We  acknowledge financial support by the European Research Council for the ERC Consolidator grant DEMOBLACK, under contract no. 770017.
 This work benefited from support by the International Space Science Institute (ISSI), Bern, Switzerland,  through its International Team programme ref. no. 393 {\it The Evolution of Rich Stellar Populations \& BH Binaries} (2017-18).

%



\software{\mobse~ \citep{2018MNRAS.474.2959G}, astropy \citep{2013A&A...558A..33A}, filltex \citep{2017JOSS....2..222G}
          }



\appendix

\section{Effects of CE efficiency on the local merger rate density} \label{sec:app}
In the main text we have assumed a fixed value for the efficiency of CE ($\alpha{}=5$). In this section, we discuss the impact of $\alpha{}$ on the merger rate density. To this purpose, we have run eight additional models varying the CE efficiency: Ej1$\alpha{}1$, Ej1$\alpha{}2$, Ej1$\alpha{}3$ and Ej1$\alpha{}4$ are the same as Ej1, but for $\alpha{}=1,$ 2, 3 and 4, respectively. Similarly, Ej2$\alpha{}1$, Ej2$\alpha{}2$, Ej2$\alpha{}3$ and Ej2$\alpha{}4$ are the same as Ej2, but for $\alpha{}=1,$ 2, 3 and 4, respectively. For each model, we have run the same set of simulations as for the ones reported in Table~\ref{tab:sims}. We find that the merger rate density of BNSs strongly correlates with the value of $\alpha{}$. Only values of $\alpha{}$ significantly larger than 2 are consistent with the BNS merger rate density inferred from the LVC. The merger rate density of BHNSs shows basically the opposite trend, with the larger value of $R_{\rm BHNS}$ being achieved for the smaller values of $\alpha{}$. Finally, the merger rate density of BBHs seems to indicate a bell-shaped dependence on $\alpha{}$, with the larger values of $R_{\rm BBH}$ obtained for $\alpha{}\sim{}2-3$. In a follow-up paper, we will discuss the physical motivations of this behavior.
\begin{figure}
  \fig{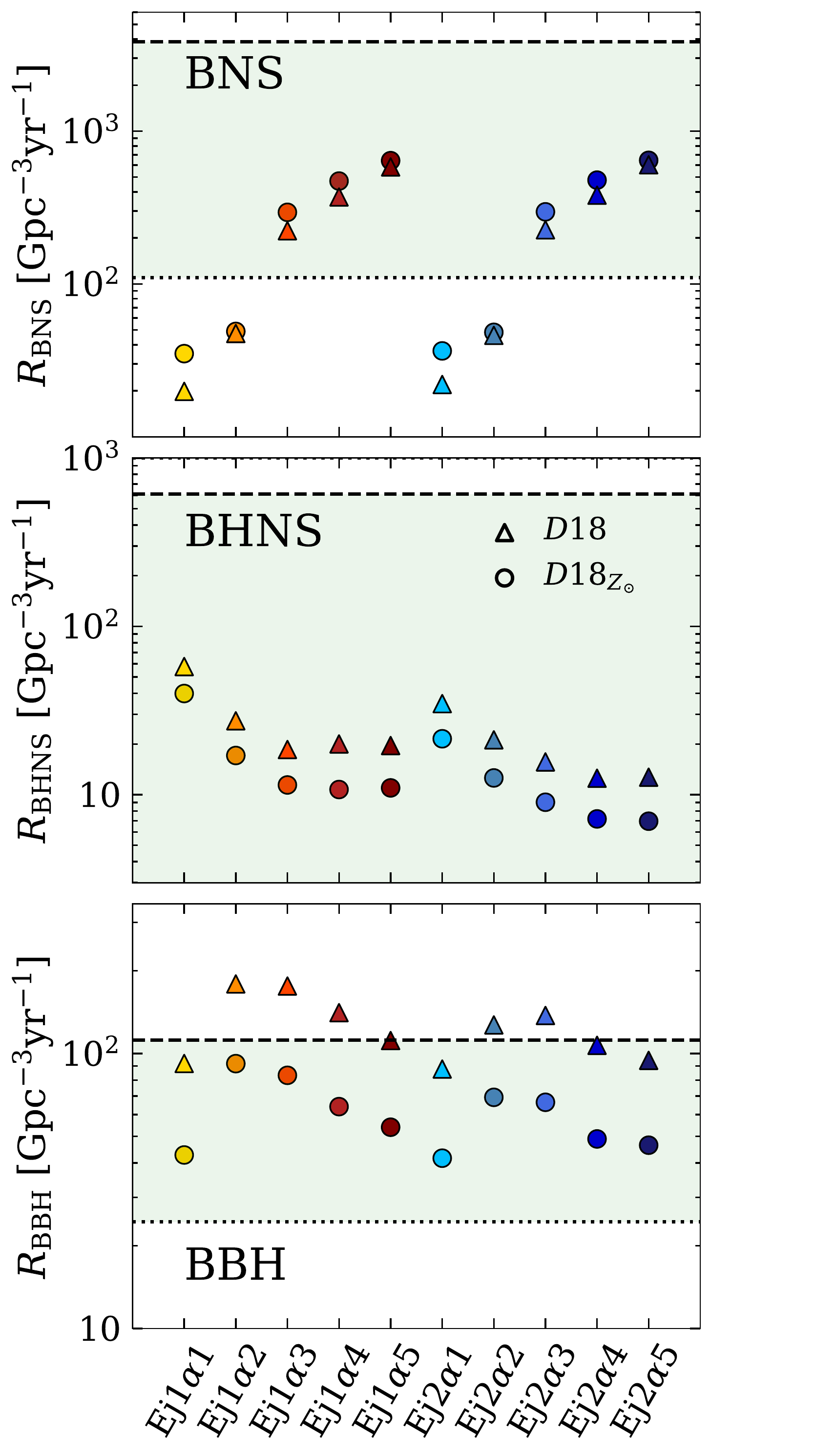}{0.47\textwidth}{}
  \caption{Local merger rate density $R$ from eq.~\ref{eq:rate}. Same as Figure~\ref{fig:Rloc}, but here we consider only models Ej1 and Ej2 and we investigate the effect of different values of CE efficiency $\alpha{}=$1, 2, 3, 4 and 5 (corresponding to models Ej1$\alpha{}1$/Ej2$\alpha{}1$, Ej1$\alpha{}2$/Ej2$\alpha{}2$, Ej1$\alpha{}3$/Ej2$\alpha{}3$, Ej1$\alpha{}4$/Ej2$\alpha{}4$ and Ej1$\alpha{}5$/Ej2$\alpha{}5$). Models labeled as Ej1$\alpha{}5$ and Ej2$\alpha{}5$ are the same as our models Ej1 and Ej2 in the rest of the paper (hence $\alpha{}=5$).}
\end{figure}




\bibliography{biblio}



\end{document}